\documentclass[1p]{elsarticle}
\usepackage{lineno}
\usepackage{xcolor}
\usepackage{amsmath}
\usepackage{graphicx}
\usepackage{caption}
\usepackage{subcaption}
\usepackage{float}

\begin{document}

\title{Performance of the Scintillation Wall in the BM@N experiment}

\author[1]{V. Volkov\corref{cor1}}
\ead{volkov@inr.ru}
\author[1]{M. Golubeva}
\author[1]{F. Guber}
\author[1]{A. Izvestnyy }
\author[1]{N. Karpushkin }
\author[1,2]{M. Mamaev}
\author[1]{A. Makhnev}

\author[1]{S. Morozov}
\author[1,2]{P. Parfenov}

\cortext[cor1]{Corresponding author}

\address[1]{Institute for Nuclear Research of the Russian Academy of Sciences,\\60-letiya Oktyabrya prospekt 7a, Moscow 117312, Russia}
\address[2]{
Joint Institute for Nuclear Research (JINR),\\Joliot-Curie 6, Dubna, Moscow region 141980, Russia}

\begin{abstract}
The performance of the scintillation wall (ScWall) has been studied in the first physics run at the Baryonic Matter at Nuclotron (BM@N) in Xe+CsI reaction at a xenon beam energy of 3.8 and 3.0 AGeV. The design and functionality of the ScWall emphasizing its ability to detect charged spectator fragments produced in nucleus-nucleus interactions are shown. The simulation results regarding ScWall's capability to determine collision geometry and the comparison between measured and simulated charged spectators fragments spectra are discussed.
\end{abstract}


\maketitle

\section{BM@N experiment}
The BM@N (Baryonic Matter at Nuclotron) experiment~\cite{Kapishin:2020cwk, Senger:2022bzm} 
is a first fixed target experiment at the NICA (Nuclotron-based Ion Collider fAcility) complex located at the Joint Institute for Nuclear Research (JINR) in Dubna, Russia. 

The goal of the experiment is to study the phase diagram of nuclear matter at high baryon densities in nucleus-nucleus collisions at ion beam energies of 2~-- 4.5 AGeV (where A is the mass number of a projectile nucleus) by measuring yields of (multi)strange hyperons, searching for hypernuclei and the azimuthal asymmetry of the charged-particle.

\begin{figure}[htbp]
  \centering
  \includegraphics[width=.7\textwidth]{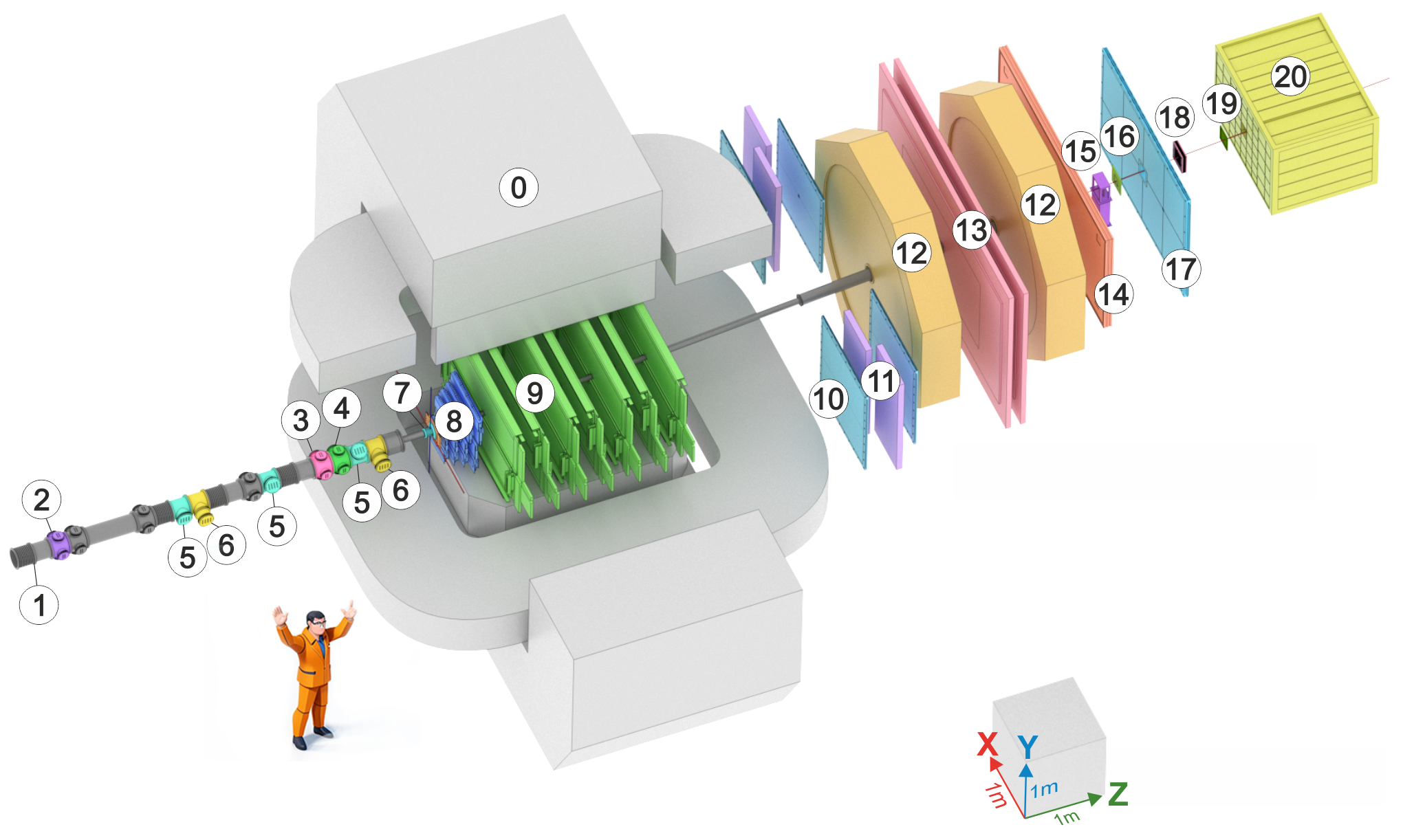}
  \caption{Schematic view of the BM@N setup in the 2023 Xe run~\cite{Afanasiev:2023opv}. Main components: 0) SP-41 analyzing magnet, 1) vacuum beam pipe, 2) BC1 beam counter, 3) Veto counter (VC), 4) BC2 beam counter, 5) Silicon Beam Tracker (SiBT), 6) Silicon beam profilometers, 7) Barrel Detector (BD) and Target station, 8) Forward Silicon Detector (FSD), 9) Gas Electron Multiplier (GEM) detectors, 10) Small cathode strip chambers (Small CSC), 11) TOF400 system, 12) drift chambers (DCH), 13) TOF700 system, 14) Scintillation Wall (ScWall), 15) Fragment Detector (FD), 16) Small GEM detector, 17) Large cathode strip chamber (Large CSC), 18) gas ionization chamber as beam profilometer, 19) Forward Quartz Hodoscope (FQH), 20) Forward Hadron Calorimeter (FHCal) .}
  \label{fig:setup}
\end{figure}

The layout of the BM@N experiment's detector system with various components for the detection of particles produced in heavy-ion collisions is shown in Figure~\ref{fig:setup}~\cite{Afanasiev:2023opv}.
The BM@N experiment consists of several detector systems:

~--- Beam Detectors~--- three silicon beam trackers determine the trajectory of beam particles upstream the target. The Beam Counter (BC2) generates a starting signal for the ToF-400 and ToF-700 detectors. 

~--- Tracking Detectors~--- the central tracking system, based on seven planes of triple gas electron multipliers (GEM), measures the momentum of charged particles in magnetic field. The system includes double-sided silicon micro-strip detectors (FSD) for high-precision interaction vertex determination and track reconstruction. The outer tracking system, with six planes of Cathode Strip Chambers (CSC), enhances track parameter precision and efficiency. 

~--- Particle Identification Detectors~--- Time-of-Flight (ToF) systems (ToF-400 and ToF-700), based on multi-gap resistive plate chamber (mRPC) technology, enable hadron ($\pi$, K, p) and light nuclei separation together with momentum up to a few GeV/c. 

~--- Collision Geometry Detectors~--- Forward Hadron Calorimeter (FHCal), Forward Quartz Hodoscope (FQH) estimate centrality and reaction plane of the collision. In addition, the ScWall is used to measure the yields of charged spectators fragments and to determine the collision geometry.

~--- Trigger System~--- The Barrel detector together with the group of beam detectors generates a trigger signal for the Data Acquisition
 (DAQ).

~--- The analyzing magnet with adjustable magnetic field (up to 1 T) optimizes detector acceptance and momentum resolution. A vacuum beam pipe minimizes multiple scattering and ionizing losses of the ion beam and produced fragments.

~--- The target station design allows switching between different target types without breaking the vacuum. 

\section{The ScWall Design and Calibration}

\begin{figure}[htbp]
  \centering
  \begin{subfigure}[b]{0.45\textwidth}
    \centering
    \includegraphics[width=\textwidth]{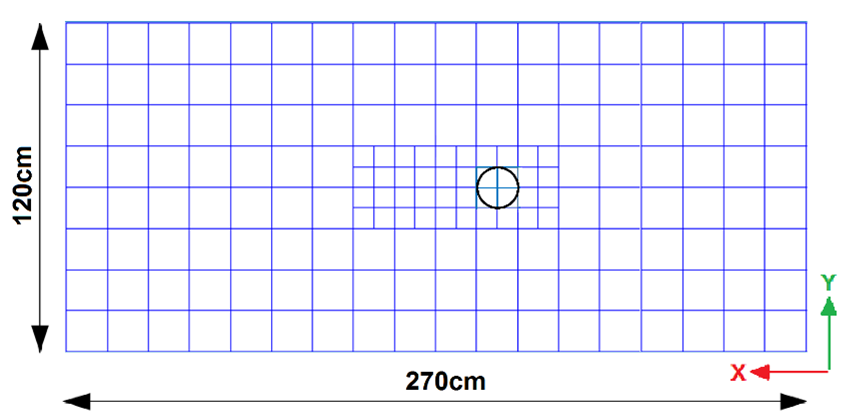}
    \caption{}
    \label{fig:scheme}
  \end{subfigure}
  \hfill
  \begin{subfigure}[b]{0.45\textwidth}
    \centering
    \includegraphics[width=\textwidth]{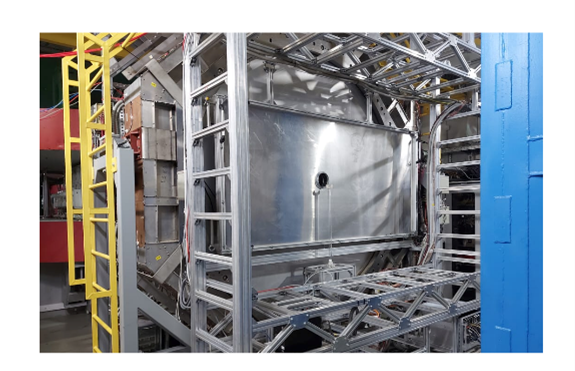}
    \caption{}
    \label{fig:overview}
  \end{subfigure}
  \caption{(a) Schematic view of the ScWall. (b) Photo of the ScWall at the BM@N setup.}
  \label{fig:scwall}
\end{figure}

The ScWall is a large-area multichannel detector designed to measure charge of detected particles in the forward rapidity region. It consists of an array of scintillator detectors housed in an aluminum box (see Figure~\ref{fig:scwall}). The ScWall comprises 40 small inner plastic scintillator detectors (\(7.5\times 7.5\times 1 \  \text{cm}^3\)) and 138 large outer detectors (\(15\times 15\times 1\ \text{cm}^3\)). To prevent radiation damage from the heavy ion beam and to reduce background in other detectors, the inner part of the ScWall has a \(15\times 15\ \text{cm}^2\) beam hole.

\sloppy
The detectors are made of polystyrene-based scintillators produced by Uniplast (Vladimir, Russia). The light generated in each scintillator is collected by WLS Y11(200) S-type (Kuraray) wavelength-shifting fibers, which is embedded in 3 mm deep grooves. At the outer fiber ends, the light is detected by Hamamatsu S13360-1325CS SiPM. This SiPM has an active area of \(1.3\times 1.3 \ \text{mm}^2\), a gain of $7\times10^5$, and a photo-detection efficiency of 25\%.

\fussy

The light yield measured with a minimum ionizing particles for the large and small detector cells is 32 and 55 photoelectrons, respectively~\cite{Baranov:2022ati}. The readout process utilizes FEE boards boards in conjunction with ADC64s2 boards. Three ADC64s2 boards~\cite{ADC64s2} digitize the signals from all ScWall detectors. More details about the ScWall design and signal readout can be found in~\cite{Volkov:2023pyd}.

\begin{figure}[htbp]
  \centering
  \begin{subfigure}[b]{0.45\textwidth}
    \centering
    \includegraphics[width=\textwidth]{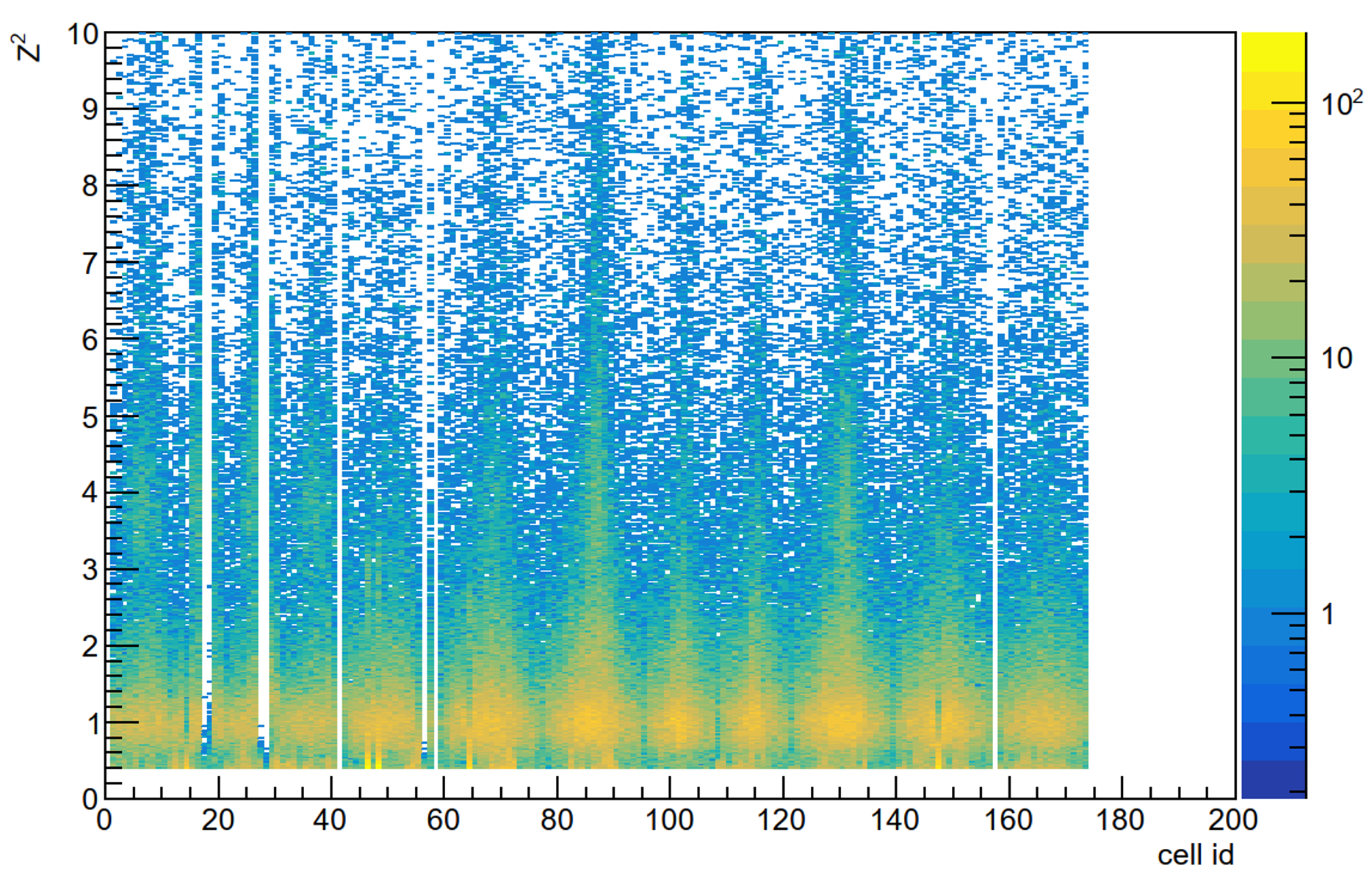}
    \caption{}
    \label{fig:spectra1}
  \end{subfigure}
  \hfill
  \begin{subfigure}[b]{0.45\textwidth}
    \centering
    \includegraphics[width=\textwidth]{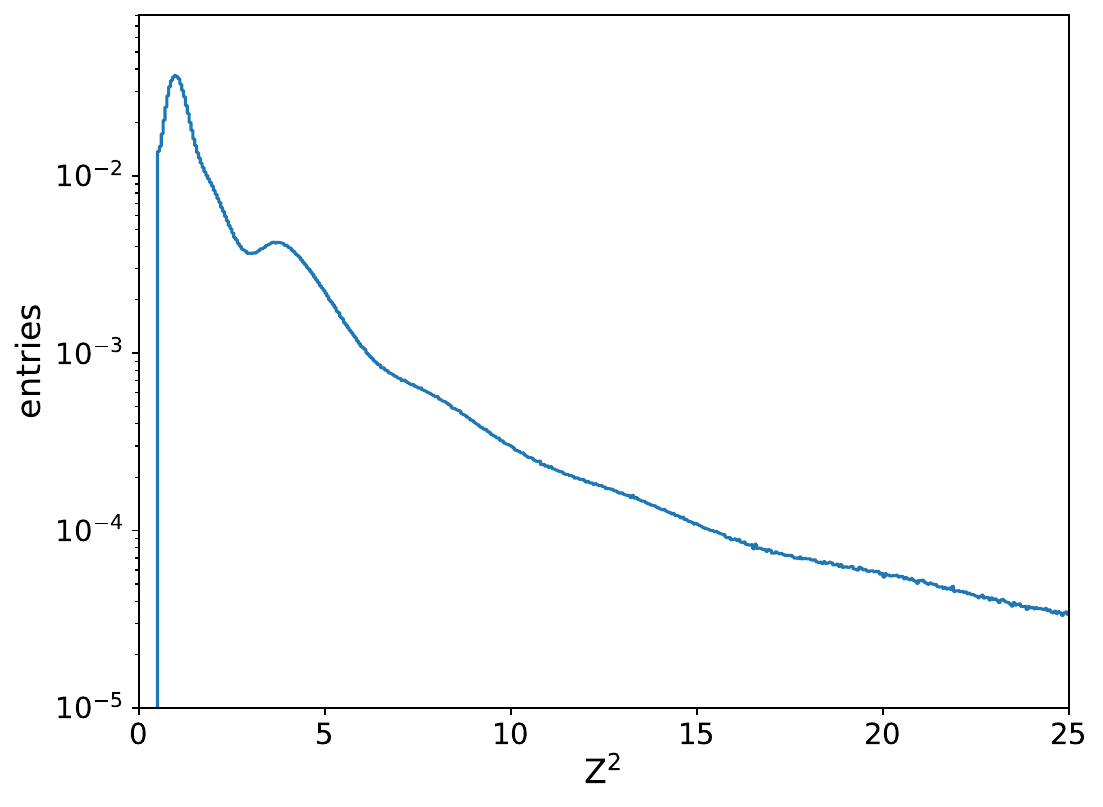}
    \caption{}
    \label{fig:spectrasmall}
  \end{subfigure}
  \hfill
  \begin{subfigure}[b]{0.45\textwidth}
    \centering
    \includegraphics[width=\textwidth]{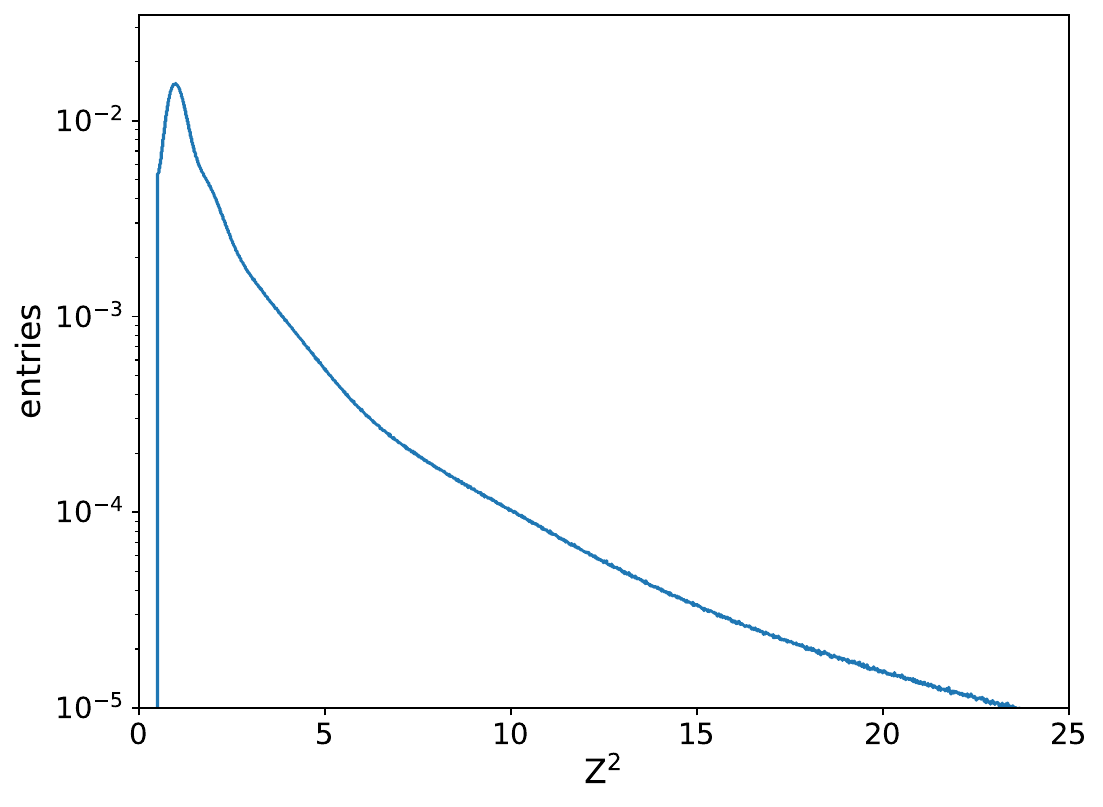}
    \caption{}
    \label{fig:spectralarge}
  \end{subfigure}
  \caption{(a) Charge spectra for all scintillator detectors in ScWall after calibration showing the prominent peak at $Z^2$ = 1. (b) Spectra of charges for small scintillator detectors after calibration. (c) Spectra of charges for large scintillator detectors after calibration.}
  \label{fig:spectra}
\end{figure}

The calibration procedure of the ScWall is aimed at equalizing signal responses of all scintillator detectors: the first peaks of the amplitude spectra corresponding to $Z$ = 1 are aligned to the same position with a calibration parameters. The result of calibration is shown in Figures~\ref{fig:spectra1} and~\ref{fig:spectrasmall}, where the first two peaks corresponding to $Z$ = 1 and $Z$ = 2 are clearly visible. The fragments with $Z$ = 3 and beyond mainly pass through the beam hole and are not detected by the most of the scintillator detectors. In the large outer scintillation detectors only the $Z$ = 1 peak being clearly visible (see Figure~\ref{fig:spectralarge}).


\begin{figure}[htbp]
  \centering
  \includegraphics[width=.7\textwidth]{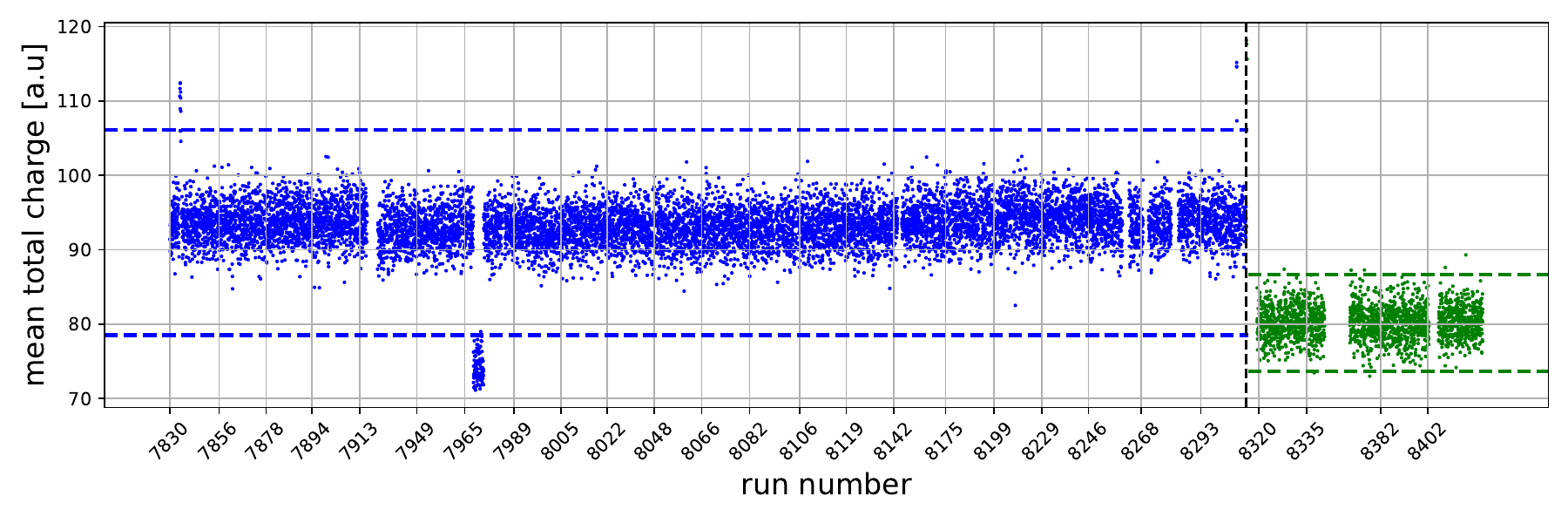}
  \caption{Mean value of the total charge on the ScWall as a function of the run number. The data for energy 3.8 AGeV are shown in blue and 3.0 AGeV in green. The dashed lines indicate the boundaries of $\pm5\sigma$.}
  \label{fig:meantotalQ}
\end{figure}

The stability of the response of the ScWall scintillator detectors during the Xe+CsI run is shown in Figure~\ref{fig:meantotalQ}. The mean value of sum of charges from all ScWall scintillator detectors remains within $\pm5\sigma$ (shown by dashed line in Figure~\ref{fig:meantotalQ}) for both 3.8 AGeV (blue dots) and 3.0 AGeV (green dots) run periods. 

\section{Comparison of the experimental charged spectra measured by ScWall with the simulated spectra }

\sloppy

\begin{figure}[htbp]
  \centering
  \includegraphics[width=.6\textwidth]{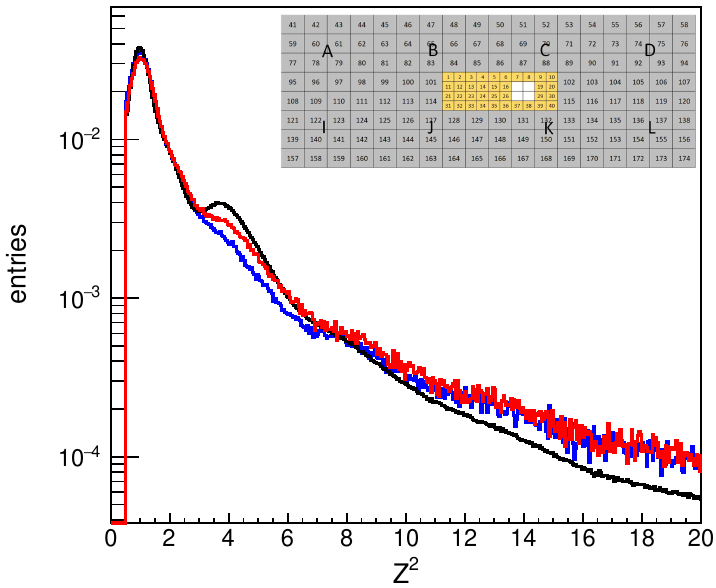}
  \caption{Experimental charge spectrum (black line) obtained in small scintillator detectors in comparison with results from PHQMD (blue line) and DCM-SMM (red line) models. The ScWall scheme highlighting the position of the small scintillator detectors is shown in insert.}
  \label{fig:expchargesp}
\end{figure}

The experimental distribution of charged fragments spectators measured by the ScWall inner part in the Xe+CsI reaction at a xenon beam energy of 3.8 AGeV for $\sim$60\% most central events is shown in Figure~\ref{fig:expchargesp} (black curve). Trigger efficiency estimation was done by comparison experimental data with realistically simulated data with DCM-QGSM-SMM model. Peaks corresponding to fragments with $Z = 1$ and $Z$ = 2 are clearly visible. There is some indication of peaks for $Z$ = 3 and $Z$ = 4, but their positions in the spectra are slightly shifted to lower values due to the Birks effect. The experimental distribution is compared with the one obtained from the DCM-QGSM-SMM ~\cite{Baznat:2019iom} and PHQMD~\cite{Aichelin:1991xy, Puri:1996qv, Dorso:1992ch, Aichelin:2019tnk} models after full chain of realistic reconstruction  (red and green curves, respectively).

\fussy

DCM-QGSM-SMM is a Monte Carlo heavy ion collision generator based on the Dubna cascade model (DCM), the quark-gluon string model (QGSM) and the statistical multifragmentation model (SMM). The model includes multiple physical phenomena: extended coalescence, multifragmentation, hyperfragment formation, and lambda polarisation.

\sloppy

The transport ‘n-body’ model PHQMD is a further development of the PHSD (Parton-Hadron-String Dynamics) model. 
PHSD includes the parton phase (quarks and gluons), the equation of state for the parton phase from lattice QCD, dynamical hadronisation and both elastic and inelastic hadron collisions in the final state. 
It also includes two-particle potential interactions between baryons as in the Quantum Molecular Dynamics (QMD) approach, where baryons are described by Gaussian-type wave functions.
PHQMD allows to choose the equation of state of matter with different compressibility modulus.

\fussy

It can be seen that both models underestimate the yield of fragments with charge $Z^2$ = 4 and overestimate yields with higher $Z^2$. In general, heavier fragments tend to be directed closer to the beam hole.

\begin{figure}[htbp]
  \centering
  \includegraphics[width=.6\textwidth]{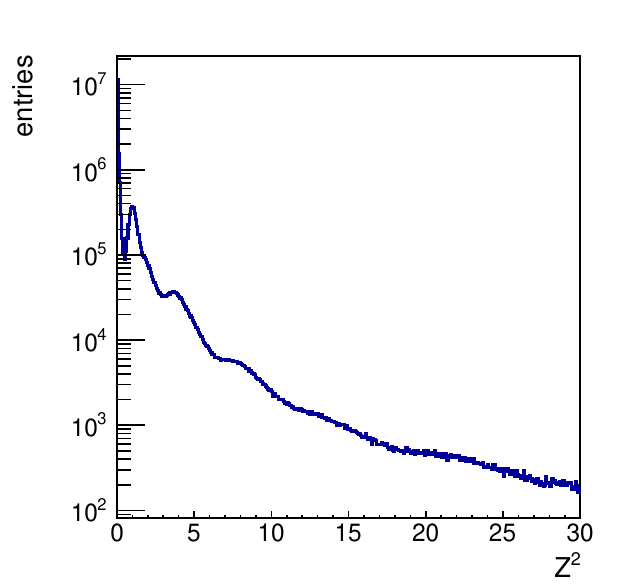}
  \caption{Experimental charge spectrum observed in one of scintillator detectors close to the beam hole exhibit a broader range of charges, up to $Z^2$ = 25.}
  \label{fig:cell9}
\end{figure}

Considering the experimental charge spectra only in scintillator detectors that are located around the beam hole, a greater range of charges is observed, up to $Z^2$ = 25 (Figure~\ref{fig:cell9}).

Thus, by comparing the obtained model distributions of charged spectator fragments with the experimental distributions, one can further tune the models to constrain the parameters of the spectator fragments production mechanisms. 


\sloppy
\section{Application of ScWall for centrality and reaction plane estimation in nucleus-nucleus collisions}
\fussy

\subsection{Centrality}
Usually, centrality is determined as a percentage of the total nuclear interaction cross-section \(\sigma\), and the centrality percentile \(c\) for an A-A collision with an impact parameter \(b\) is determined by integrating the distribution of the impact parameter:
\begin{equation}
c = \frac{\int_0^b \frac{d\sigma}{db'} db'}{\int_0^\infty \frac{d\sigma}{db'} db'} = \frac{1}{\sigma_{AA}} \int_0^b \frac{d\sigma}{db'} db'.
\end{equation}
Generally, there are several observables that are sensitive to the initial event geometry such as charged particles multiplicity~\cite{STAR:2018zdy, HADES:2017def}, as well as energy of spectators in forward rapidity region~\cite{NA61SHINE:2021nye, Guber:2022pcj}. The total charge of spectators measured by scintillator detectors in the inner part of ScWall can be used for this purpose as well.

\begin{figure}[htbp]
  \centering
  \begin{subfigure}[b]{0.45\textwidth}
    \centering
    \includegraphics[width=\textwidth]{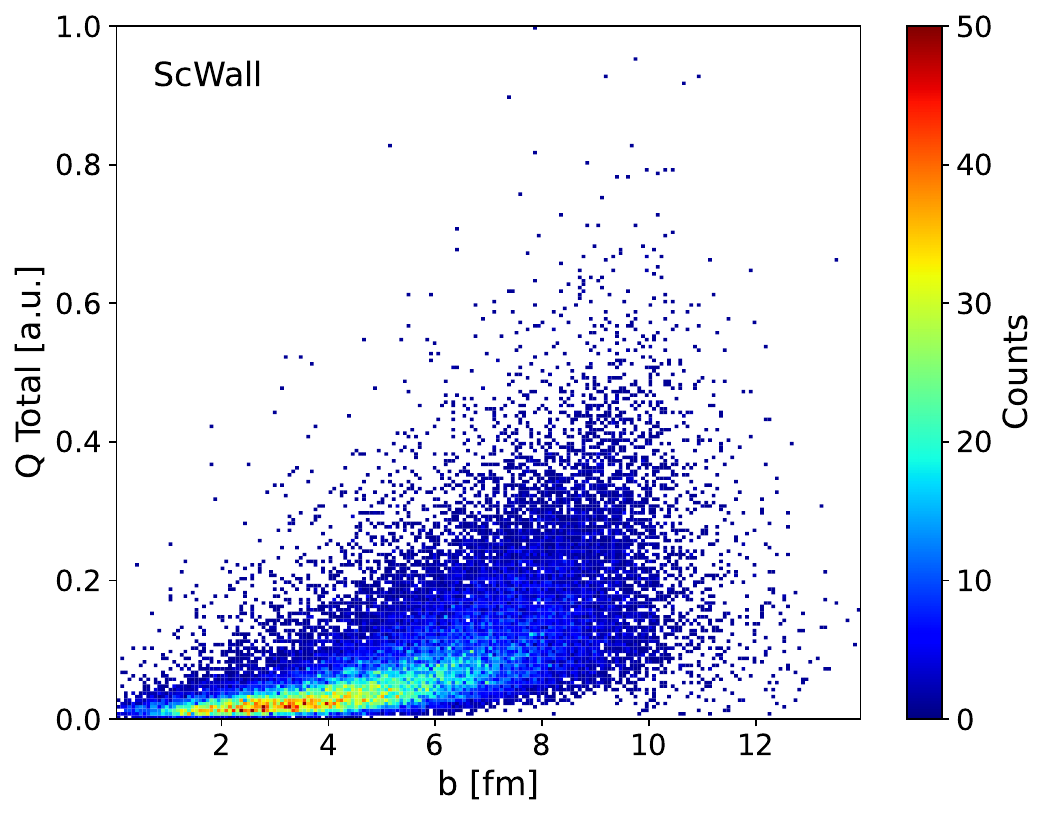}
    \caption{}
    \label{fig:scwallb}
  \end{subfigure}
  \hfill
  \begin{subfigure}[b]{0.45\textwidth}
    \centering
    \includegraphics[width=\textwidth]{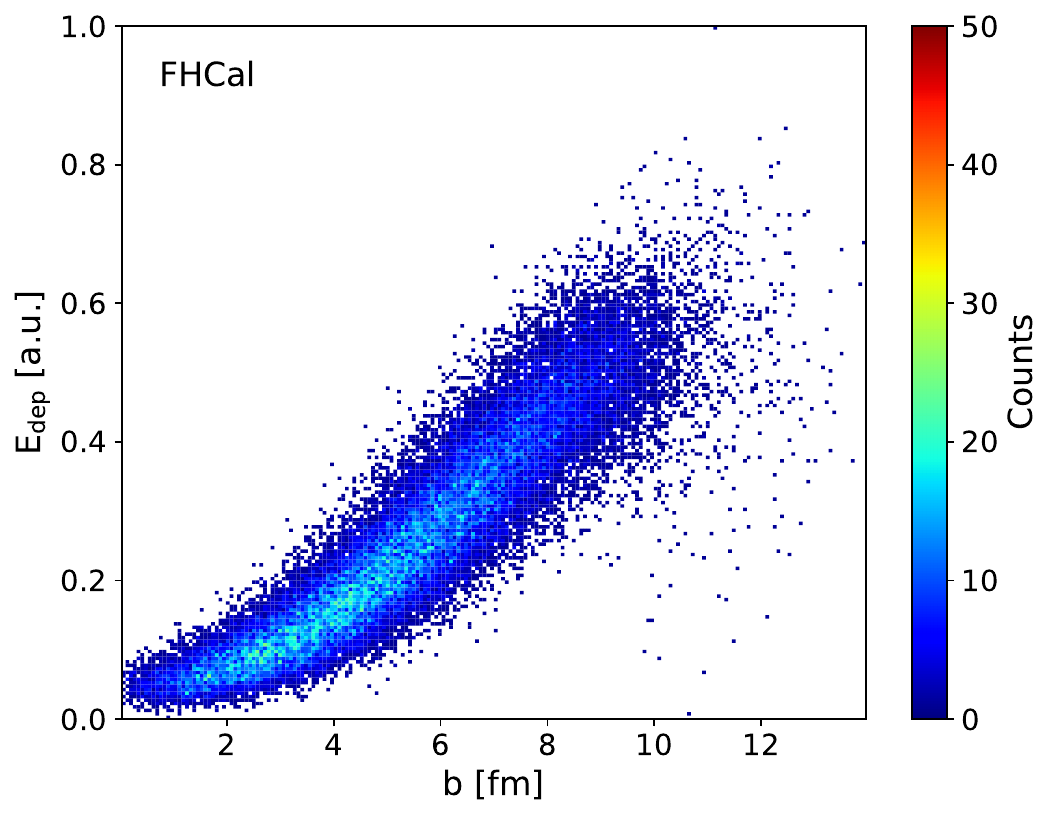}
    \caption{}
    \label{fig:fhcalb}
  \end{subfigure}
  \caption{(a) Correlation between total charge in ScWall and impact parameter. (b) Correlation between deposited energy in FHCal and impact parameter $b$. Both distributions are obtained from the DCM-QGSM-SMM model for 60\% of the  most central collisions.}
  \label{fig:corr}
\end{figure}

Centrality classes were determined using a straightforward method: dividing the one-dimensional distribution of the studied observable into bins with an equal number of events.
To address the ambiguity in the dependence of total charge in ScWall and deposited energy in FHCal on centrality, due to beam holes in both detectors, additional information is incorporated. A criterion on the total charge measured in the FQH, placed in front of the FHCal beam hole, was chosen~\cite{Guber:2021exv}. This criterion effectively selects approximately 60\% of the most central events. The observed correlation between the total charge in ScWall and the impact parameter $b$ simulated using the DCM-QGSM-SMM model (Figure~\ref{fig:scwallb}) is not so strong as the correlation between the deposited energy in the FHCal and $b$ (Figure~\ref{fig:fhcalb}). This difference is mainly due to the fact that FHCal measures the energy of charged spectators as well as neutrons.





\begin{figure}[htbp]
  \centering
  \begin{subfigure}[b]{0.45\textwidth}
    \centering
    \includegraphics[width=\textwidth]{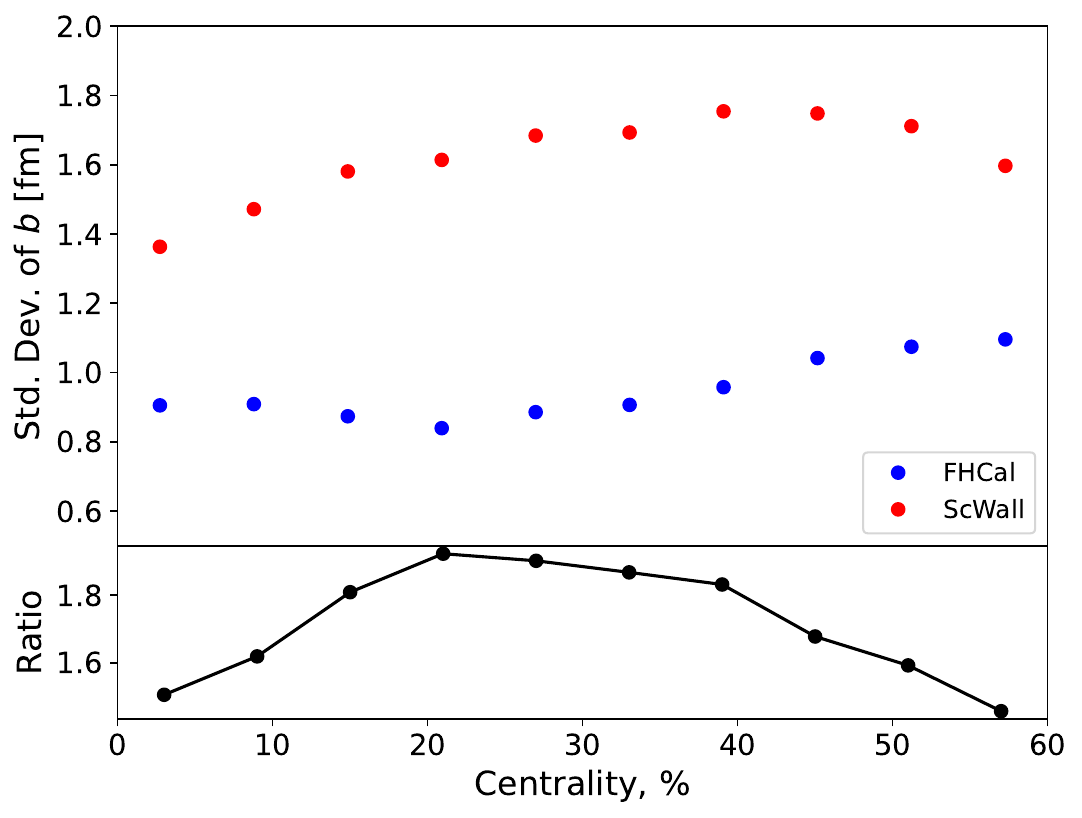}
    \caption{}
    \label{fig:stddev}
  \end{subfigure}
  \hfill
  \begin{subfigure}[b]{0.45\textwidth}
    \centering
    \includegraphics[width=\textwidth]{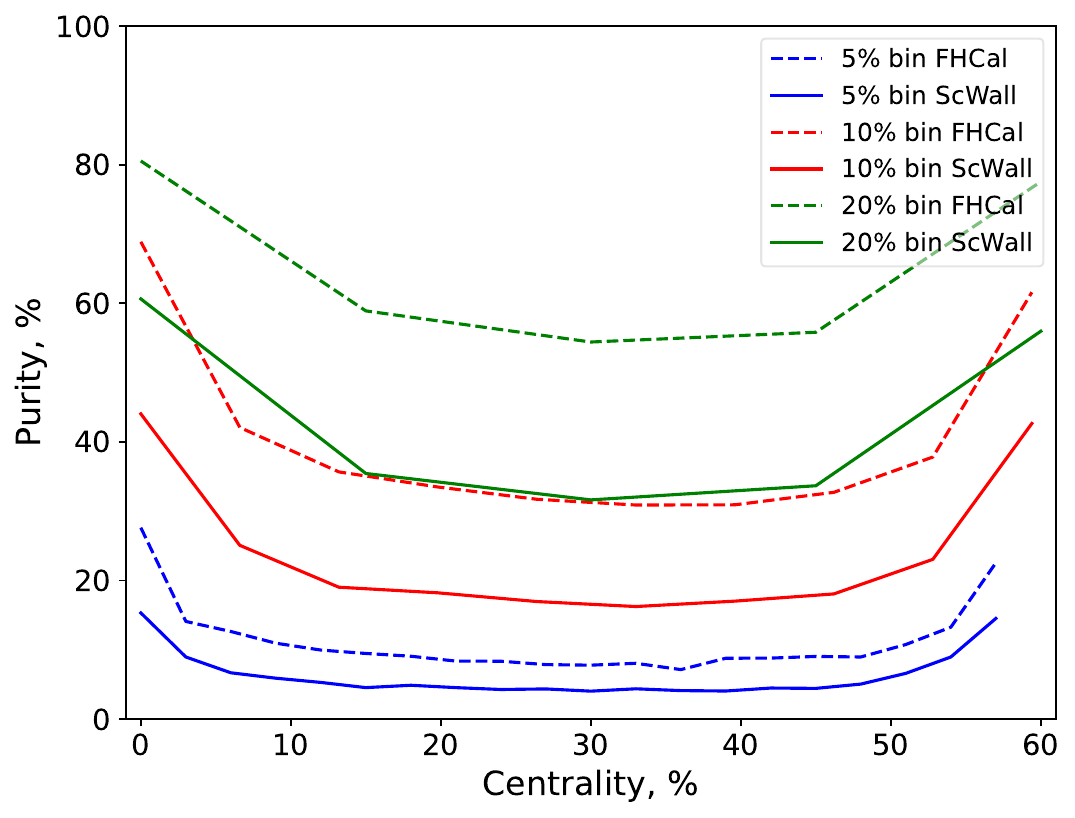}
    \caption{}
    \label{fig:purity}
  \end{subfigure}
  \caption{(a) The standard deviation of impact parameter distributions in 10 centrality classes plotted against centrality classes using different observables: ScWall total charge distribution  (red) and FHCal total deposited energy distribution  (blue). The lower panel shows ScWall/FHCal observables  ratio. (b) Purity as function of centrality classes obtained from ScWall (solid lines) and FHCal (dashed lines) observables for different centrality class width: 5\%, 10\%, 20\%.}
  \label{fig:stddevb2}
\end{figure}

As seen in Fig. 7, the standard deviation of the impact parameter distribution across 10 classes, plotted against centrality is shown in Figure~\ref{fig:stddev} and demonstrates that the ScWall-based centrality classes exhibit a significantly broader spread compared to those derived from the FHCal. Thus, the highest precision in centrality determination in the BM@N is provided by the tracks multiplicity~\cite{Segal:2023njv}, or by the FHCal energy deposition while the ScWall exhibits lower sensitivity to centrality. However, the ScWall can provide valuable systematic information when used in conjunction with the FHCal and FQH.

Certain measurements in heavy-ion physics~\cite{HADES:2020wpc} are highly sensitive to the purity of centrality classes. The purity parameter, $p$, is defined as $p = 1 - N_{overlap}/N_{class}$, where $N_{overlap}$ represents the number of events shared between centrality classes, and $N_{class}$ is the total number of events in the classes.
 From Figure~\ref{fig:purity} it can be seen that to achieve a maximum purity value of 80\% the width of the the most central class should be not less than 20\%. This finding shows the importance of centrality class width choice in maintaining high purity in centrality measurements, providing a critical benchmark for future analyses. Here again better purity is observed if the FHCal energy deposition is used for the centrality classification.


\subsection{Comparison of reaction plane determination by ScWall and FHCal}

One of the most significant geometry parameter in relativistic heavy ion collisions is the reaction plane of the collision $\Psi_{RP}$ which is spanned on the vectors of impact parameter and beam direction. 
Relative to this reaction plane the azimuthal flow of the produced particles, global polarization and many other collective effects sensitive to the transport and bulk parameters of the matter created in the collision are calculated. 

Every quantity sensitive to collective effects (i.e. global polarization, or collective flow, etc.) measured relative to the estimated event plane $\Psi_{EP}$ will have the systematic bias compared to same quantity measured with respect to true reaction plane $\Psi_{RP}$ since the estimated $\Psi_{EP}$ fluctuates relative to the true $\Psi_{RP}$. 
The measured quantity with respect to the $\Psi_{EP}$ is divided over the resolution correction coefficient as for example for the azimuthal flow of the produced in the collision particles~\cite{Voloshin:1998} or in the case with anizotropic flow of particles:

\begin{equation}
    v_n = \frac{ v_n\{\Psi_{EP} \} }{ R_n },
\end{equation}

where $v_n\{\Psi_{EP}\}$ is the $n^{th}$ harmonic flow estimation relative the $\Psi_{EP}$. The absolute statistical error on the measured directed flow can be expressed as follows:
\begin{equation}
\Delta v_n = \sqrt{\left(\frac{\Delta v_n \ \Psi_{EP}}{R_n}\right)^2 + \left(\frac{v_n \ \Psi_{EP}}{R_n^2} \Delta R_n \right)^2}
\end{equation}

From this equation we see that the error on the measured directed flow is inversely proportional to the symmetry plane resolution correction factor. The greater the resolution correction coefficient of the symmetry plane the less statistics are needed to satisfy the required accuracy level of the $v_n$.

\begin{figure}[htbp]
  \centering
  \includegraphics[width=.8\textwidth]{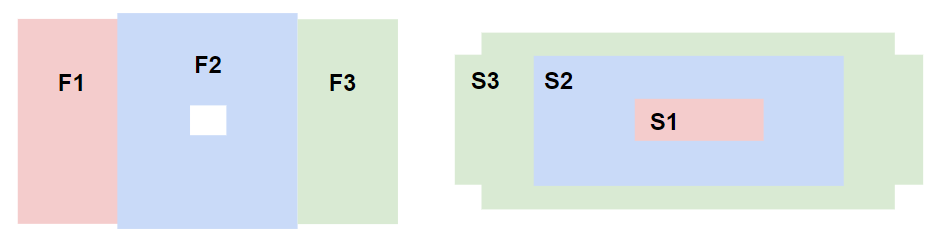}
  \caption{Schematic representation of the groups of modules of FHCal (left) and ScWall (right) used to estimate the event plane.}
  \label{fig:groups}
\end{figure}

In the BM@N experiment the event plane is estimated using the azimuthal asymmetry in the distribution of spectator particles which are deflected in the reaction plane by the expanding hot and dense matter within the overlap region. Two detector subsystems can measure this azimuthal distribution of spectators, which are the FHCal~\cite{Mamaev:2023fpr} and ScWall. Modules of the FHCal or scintillator detectors of the ScWall were divided into 3 groups and the event plane in each group was estimated separately. Areas shown on the Figure~\ref{fig:groups} by different colors is the schematic representation of the groups of modules of FHCal and ScWall used to estimate the event plane.

\begin{figure}[htbp]
  \centering
  \includegraphics[width=.6\textwidth]{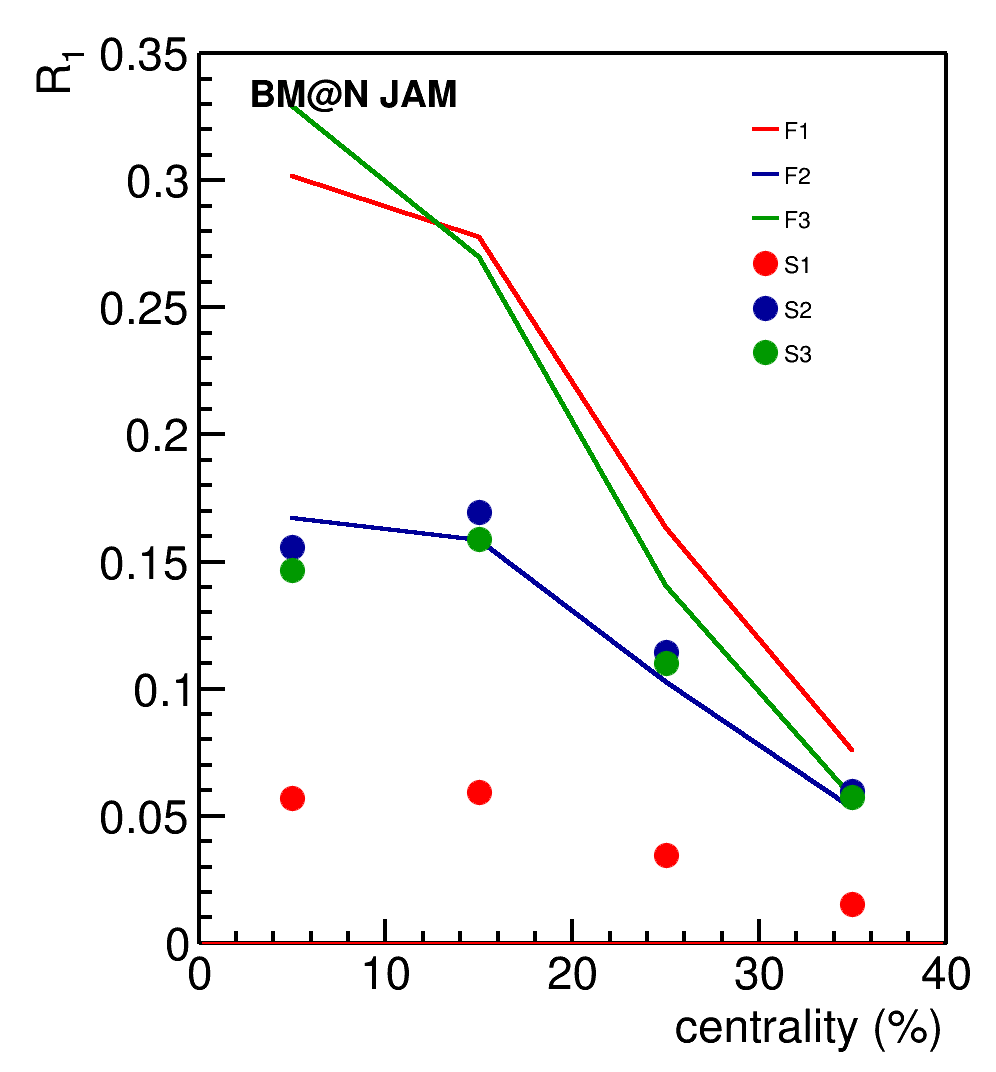}
  \caption{Event plane resolution R1 as a function of centrality for different groups of modules of FHCal and ScWall calculated in JAM model.}
  \label{fig:EP}
\end{figure}

The simulation of event plane resolution correction coefficient $R_1$ as a function of centrality for different groups of modules of FHCal and ScWall is presented on the Figure~\ref{fig:EP}.
Collisions generated by the Jet A-A Model (JAM)~\cite{Nara:2019crj} model with momentum-dependent potential were used as an input for the simulation.
Then, these events were propagated through the BM@N detector medium using the GEANT4 transport code and then a full chain of realistic reconstruction was employed.
The JAM model was used since it provides the reliable flow signal for baryons which is crucial in studying the performance for the symmetry plane reconstruction.
Event plane resolution coefficient $R_1$ for the groups of modules of FHCal is significantly exceeds the one observed for ScWall. Therefore, based on this we can conclude less statistics is needed to estimate the collective flow of the produced particles measured with respect to FHCal symmetry plane. Nevertheless, the detector subsystem ScWall is still suitable for the event plane estimation and can be utilized for systematics evaluation. For example, the collective flow can be measured relative to the FHCal and ScWall symmetry planes and results can be compared to estimate the error due to symmetry plane estimation.

\section{Conclusion}
The performance of the ScWall in the first physics run at BM@N experiment is presented. In the Xe+CsI reaction at 3.8 AGeV, fragments with charges $Z$ = 1 and $Z$ = 2 are observed in the central small scintillator detectors, while fragments with charges up to $Z$ = 5 are detected only in the detectors closest to the beam hole. The DCM–QGSM–SMM and PHQMD describe the yields of fragments with $Z$ = 1 rather good, however an overestimation of the yields for higher Z is presented, indicating potential limitations in their predictive accuracy for fragmentation processes. This study demonstrates the need to adjust fragmentation mechanisms within these models to better describe the production of spectator fragments in nucleus-nucleus collisions. Obtained experimental data from the ScWall at the BM@N experiment offers valuable insights for optimizing model parameters in this energy range.

Centrality estimation using the ScWall's total charge shows less correlation with impact parameter b in comparison with impact parameter correlation with energy deposition in the FHCal, indicating superior performance of the latter in determining centrality. However, ScWall can be used to estimate systematics of centrality determination. 

Purity analysis reveals that achieving 80\% purity in the most central classes requires defining centrality class sizes not less than 20\% by energy deposition in the FHCal, emphasizing the importance of careful class definition for accurate centrality measurements. 

It is shown that the ScWall and FHCal detectors effectively estimate the event plane in heavy-ion collisions at the BM@N with FHCal exhibiting better resolution correction coefficient than the ScWall making it preferable for precise flow measurements.

\section*{Acknowledgments}
The work was supported by the Ministry of Science and Higher Education of the Russian Federation, Project FFWS-2024-0003.

\bibliographystyle{elsarticle-num}

\end{document}